\documentclass[11pt,a4paper]{article}
\usepackage{jheppub}

\usepackage{amsfonts}
\usepackage{slashed}

\def\tr{{\rm Tr}}
\def\det{{\rm det\,}}

\def\bea{\begin{eqnarray}}
\def\eea{\end{eqnarray}}
\def\nn{\nonumber}
\def\half{\frac{1}{2}}

\def\dash{\,\textendash\, }
\def\lmatrix{\left(\begin{array}}
\def\rmatrix{\end{array}\right)}
\def\mysum{\sum_{n\neq 0}}
\def\eps{\varepsilon}
\def\msbar{\overline{\rm MS\kern-0.5pt}\kern0.5pt}

\title{The Yang-Mills gradient flow in finite volume}

\author[abc]{Zoltan Fodor,}
\author[de]{Kieran Holland,}
\author[f]{Julius Kuti,}
\author[c]{Daniel Nogradi}
\author[f]{and Chik Him Wong}

\affiliation[a]{University of Wuppertal, Department of Physics, Wuppertal D-42097, Germany}
\affiliation[b]{J\"ulich Supercomputing Center, Forschungszentrum J\"ulich, J\"ulich D-52425, Germany}
\affiliation[c]{E\"otv\"os University, Institute for Theoretical Physics, Budapest 1117, Hungary}
\affiliation[d]{University of the Pacific, 3601 Pacific Ave, Stockton CA 95211, USA}
\affiliation[e]{Institute for Theoretical Physics, Albert Einstein Center for Fundamental Physics, Bern University, Sidlerstrasse 5, CH-3012 Bern, Switzerland}
\affiliation[f]{University of California, San Diego, 9500 Gilman Drive, La Jolla, CA 92093, USA}

\emailAdd{fodor@bodri.elte.hu}
\emailAdd{kholland@pacific.edu}
\emailAdd{jkuti@ucsd.edu}
\emailAdd{nogradi@bodri.elte.hu}
\emailAdd{rickywong@physics.ucsd.edu}

\abstract{The Yang-Mills gradient flow is considered on the four dimensional torus $T^4$ for $SU(N)$ gauge
theory coupled to $N_f$ flavors of massless fermions in arbitrary representations. The small volume dynamics is dominated by the constant gauge fields.
The expectation value of the field strength tensor
squared $\tr F_{\mu\nu} F_{\mu\nu}(t)$ is calculated for positive flow time $t$ by treating the non-zero gauge modes
perturbatively and the zero modes exactly. The finite volume correction to the infinite volume result is found to contain both
algebraic and exponential terms. The leading order result is then used
to define a one parameter family of running coupling schemes in which the coupling runs with the linear size of the box.
The new scheme is tested numerically in $SU(3)$ gauge theory coupled to $N_f = 4$ flavors of massless fundamental fermions. 
The calculations are performed at
several lattice spacings with a controlled continuum extrapolation. The continuum result agrees with the perturbative
prediction for small renormalized coupling as expected.}

\arxivnumber{1208.1051}

\begin{document}

\maketitle

\section{Introduction and summary}
\label{intro}

The Yang-Mills gradient flow \dash or Wilson flow \dash has proved to be a useful tool in lattice gauge theory. 
In the context of the Nicolai map it was studied in \cite{Luscher:2009eq}; see also \cite{Narayanan:2006rf}
for an earlier appearance. A systematic investigation, including suggestions for possible applications, has 
appeared relatively recently \cite{Luscher:2010iy, Luscher:2010we, Luscher:2011bx}. See also \cite{Lohmayer:2011si}. The
first concrete very useful application of the flow for high precision setting of the physical scale in QCD simulations has been presented in
\cite{Borsanyi:2012zs}. The flow in QCD applications has so far been considered in infinite volume which is most appropriate for low
energies.

In the present work the flow is calculated on the four dimensional torus, i.e. in a finite four dimensional box. The motivation
for doing so is to obtain a new running coupling scheme in which the renormalized coupling runs with the linear size of the box. In principle the
original infinite volume flow can also be used for defining a renormalized running coupling $g_R(q)$ with $q = 1 / \sqrt{8t}$
where $t$ is the flow time, but the control of finite volume corrections is an additional concern in this case. 
This issue is eliminated if the running $g_R(L)$ is with the linear size $L$. In particular, a step scaling analysis can be
performed \cite{Luscher:1991wu, Luscher:1992an}.

Due to asymptotic freedom perturbation theory is reliable for small volumes hence the appropriate framework is the small volume
expansion or femto world \cite{Luscher:1982ma, Koller:1985mb, Koller:1987fq, vanBaal:1988va, vanBaal:1988qm}; see also 
\cite{KorthalsAltes:1985tv, Coste:1985mn, Coste:1986cb, KorthalsAltes:1988is}.
The usual complication associated with calculations in the femto world is the presence of gauge zero
modes which dominate the dynamics and are not Gaussian. They need to be treated exactly while the gauge non-zero modes can be
integrated out in perturbation theory. As will be shown, the contribution of the non-zero modes renormalizes the bare coupling
according to the 1-loop $\beta$-function and generates an effective action for the zero modes.

The quantity which turns out to be the most useful for our purposes is the one that has already been calculated in infinite volume
in \cite{Luscher:2010iy}, namely the field strength squared at $t>0$ flow time,
\bea
\label{e}
E(t) = -\frac{1}{2} \tr F_{\mu\nu} F_{\mu\nu}(t)
\eea
(see appendix \ref{conventions} for our conventions). The expansion of its expectation value in finite volume 
is our main result and to leading order in the $\msbar$ scheme it is given by
\bea
\label{run}
\langle t^2 E(t) \rangle = g_R^2(\mu) \frac{3(N^2-1)}{128\pi^2}\left( 1 + \delta \right)
\eea
where $\mu$ is the dimensional regularization scale, $g_R^2(\mu)$ is the renormalized coupling in the $\msbar$ scheme.
The correction factor $\delta = \delta_a + \delta_e$ is a sum of algebraic and exponential terms,
\bea
\delta_a &=& - \frac{64t^2\pi^2}{3L^4} \nn \\
\delta_e &=& \vartheta^4\left(\exp\left(-\frac{L^2}{8t}\right)\right) - 1 = 8 \exp\left(-\frac{L^2}{8t}\right) + 24
\exp\left(-\frac{L^2}{4t}\right) + \ldots\;,
\eea
and where $\vartheta(q)$ is the standard Jacobi elliptic function (normally called $\vartheta_3(q)$, see appendix
\ref{conventions} for details). Indeed, the infinite volume result in \cite{Luscher:2010iy} is reproduced.

Equation (\ref{run}) can be used to define a running coupling $g_R(L)$
which will run with the linear size once the dimensionless combination $c = \sqrt{8t}/L$ is held fixed and $\mu = 1/L$ is set. 
Different choices for $c$ correspond to different schemes.

The organization of the paper is as follows. In section \ref{smallvolume} the small volume expansion is given on $T^4$ and the
finite effective action for the gauge zero modes is calculated by integrating out the non-zero modes to 1-loop. In order for the
presentation to be self-contained all details are spelled out although the methods are by no means new. In section \ref{ymgrad} the
gradient flow is considered and the expectation value of the quantity $E(t)$ is calculated by again treating the non-zero modes in
1-loop perturbation theory and using the previously obtained effective action for the zero modes. The result is then used in 
section \ref{runningcoupling} to define a renormalization scheme for the gauge coupling. As an illustration of the method,
numerical simulations are used to compute the running coupling in $SU(3)$ gauge theory coupled to $N_f = 4$ massless quarks in
section \ref{numericalresults}. Finally we close with conclusions and provide an outlook in section \ref{conclusion}.

\section{Small volume expansion}
\label{smallvolume}

On the four dimensional Euclidean torus $T^4$ with periodic boundary conditions for the gauge field the zero momentum (constant)
gauge mode is separated from the first non-zero momentum mode by the gap $2\pi/L$ and dominates the low energy small volume
dynamics \cite{Luscher:1982ma}; see also \cite{Koller:1985mb, Koller:1987fq, vanBaal:1988va, vanBaal:1988qm, KorthalsAltes:1985tv, Coste:1985mn, Coste:1986cb, KorthalsAltes:1988is}.
This dynamics is non-linear because of the quartic interaction and needs to be treated exactly while the dynamics of the non-zero modes can be treated
perturbatively. Correspondingly the gauge field is split
\bea
\label{abq}
A_\mu(x) = B_\mu + Q_\mu(x)\;,\qquad \int d^4x Q_\mu(x) = 0
\eea
into the zero mode $B_\mu$ and non-zero modes $Q_\mu(x)$. The action for $N_f$ flavors of massless Dirac fermions in
representation $R$ is
\bea
S = - \frac{1}{2g_0^2} \int d^4 x \tr F_{\mu\nu} F_{\mu\nu} + \sum_{f=1}^{N_f} \int d^4 x {\bar\psi_f} \slashed{D} \psi_f\;.
\eea
where $g_0$ is the bare coupling constant.
The boundary condition for the fermions is assumed to be anti-periodic in at least one direction. 
It is convenient to introduce $\partial_\mu + B_\mu =
D_\mu(B)$ acting in either the adjoint or representation $R$ depending on whether it is applied to a gauge field or fermion.

Gauge fixing is only required for the gauge non-zero modes and 
a convenient gauge choice is the background gauge $\chi = D_\mu(B) Q_\mu = 0$.
The constant gauge transformations do not need to be fixed as their volume is finite.

Neglecting interactions which are higher order in $Q_\mu$ and the ghost field one obtains the leading order Faddeev-Popov
operator as $D_\mu(B)^2$
which is understood in the adjoint representation and acts on ghosts without zero-modes. The corresponding effective action for
the zero mode $B_\mu$ is then 
\bea
S_{gh}(B) = - \ln\det( D_\mu(B)^2 )\;.
\eea

The quadratic term in $Q_\mu$ from the gauge action is
\bea
\frac{1}{2g_0^2} \int d^4x \tr Q_\mu \left( D_\rho(B)^2 \delta_{\mu\nu} - D_\mu(B) D_\nu(B) + 2 [B_\mu,B_\nu] \right) Q_\nu\;.
\eea
A convenient way of implementing gauge fixing is by adding $\chi^2/2g_0^2$ to the action which allows integrating out
the $Q_\mu$ field without the gauge constraint. The effective action from this bosonic integral is then,
\bea
S_{Q}(B) = \half\ln\det\left( D_\rho(B)^2 \delta_{\mu\nu} + 2 [B_\mu,B_\nu] \right)\;.
\eea

In the fermionic action one may neglect the interaction between the $Q_\mu$ fields and the fermions. To leading order one
obtains the effective action
\bea
S_{F}(B) = - \ln \det( \slashed{D}(B) )^{N_f} = - \ln \det\left( D_\mu(B)^2 + \half\sigma_{\mu\nu} [B_\mu, B_\nu] \right)^{N_f/2}\;,
\eea
where $\sigma_{\mu\nu} = [\gamma_\mu,\gamma_\nu] / 2$.
Here the operators act on fermions with the appropriate boundary condition. The various determinants will be evaluated using
dimensional regularization and all subsequent calculations are done in dimension $d = 4 - 2\eps$.

The total effective action after integrating out the gauge non-zero modes, the ghosts and the fermions is then
\bea
\label{seff}
S_{eff}(B) = -\frac{L^4}{2g_0^2(\mu L)^{2\eps}}\tr [B_\mu,B_\nu]^2 + S_Q(B) + S_{gh}(B) + S_F(B)\;,
\eea
where the first term is the tree level action for the constant mode and $\mu$ is the scale of dimensional
regularization.

Now we will proceed to evaluating the various determinants. They will be Taylor-expanded in $B_\mu$ and we will see later that it is
enough to expand them to fourth order for our purposes. Higher orders in $B_\mu$ will correspond to higher orders in the
renormalized coupling. The expansion is around the free $B_\mu = 0$ determinants and these (infinite) constants are dropped as
usual.

The derivatives in $S_Q$ and $S_{gh}$ are replaced by $2\pi i n_\mu / L$ where $n_\mu$ are integers and $n^2 \neq 0$. In $S_F$ the
derivatives are replaced by $2\pi i ( n_\mu - k_\mu ) / L$ where $k_\mu$ is $1/2$ in all anti-periodic fermion directions and
the rest of its components are zero. We will assume $k^2 \neq 0$. 
It is furthermore convenient to introduce the hermitian matrices $C_\mu = L B_\mu / 2\pi i$.

Straightforward calculation yields that up to fourth order in $C_\mu$ the following holds
\bea
\label{egy}
S_Q(C) + S_{gh}(C) = \tr_{ad}\log( D_\mu(C)^2 ) + \gamma\, \tr_{ad} [ C_\mu, C_\nu ]^2\;,
\eea
where the traces are in the adjoint representation and
\bea
\gamma = \sum_{n\neq 0} \frac{1}{n^4}\;.
\eea
Similarly, the fermionic contribution to the effective action up to fourth order in $C_\mu$ is
\bea
\label{ketto}
S_F(C) = -2N_f \left( \tr_R\log( D_\mu(C)^2 ) + \frac{\gamma(k)}{4} \tr_R [C_\mu,C_\nu]^2 \right)\;,
\eea
where all traces are in the representation $R$ and
\bea
\gamma(k) = \sum_n \frac{1}{(n-k)^4}\;.
\eea
Equations (\ref{egy}) and (\ref{ketto}) show that only the Laplacian is needed in the background of $C_\mu$ in arbitrary
representation and with arbitrary boundary condition in order to evaluate the full effective action.

First, let us evaluate all determinants with periodic boundary condition and get back to the case of non-trivial boundary
conditions for the fermions later. Explicit calculation yields up to fourth order in $C_\mu$,
\bea
\label{laplace1}
-\tr_R\log( D_\mu(C)^2 ) &=& \delta\, \frac{2-d}{d} \, \tr_R C^2 + \gamma\, \frac{d-8}{2d}\, \tr_R C^4 + \nn \\
&&+ \, 4 \, \mysum \frac{ n_\mu n_\nu n_\rho n_\sigma }{n^8} \tr_R C_\mu C_\nu C_\rho C_\sigma\;,
\eea
where the new constant $\delta$ has been introduced and $C^2 = C_\mu C_\mu$ and $C^4 = (C_\mu C_\mu)^2$ are $SO(4)$ invariant
combinations. It is useful to define two more constants $\alpha$ and $\beta$ by
\bea
\delta = \mysum \frac{1}{n^2}\;,\qquad \alpha = \mysum \frac{n_1^4}{n^8}\;,\qquad \beta = \mysum \frac{n_1^2 n_2^2}{n^8}\;.
\eea
Using these the following is easy to show,
\bea
\label{cnegy}
\mysum \frac{ n_\mu n_\nu n_\rho n_\sigma }{n^8} \tr_R C_\mu C_\nu C_\rho C_\sigma = (\alpha - 3\beta) \sum_\mu C_\mu^4 + 
\beta \left( 3 \tr_R C^4 + \frac{1}{2} \tr_R [C_\mu,C_\nu]^2 \right).\;\;\;\;\;\;
\eea
Since the torus breaks rotations the $SO(4)$-breaking first term on the right hand side is allowed. Combining equations
(\ref{laplace1}) and (\ref{cnegy}) we obtain,
\bea
\label{abc}
-\tr_R\log( D_\mu(C)^2 ) &=& \delta \, \frac{\eps-1}{2-\eps} \tr_R C^2 + 4(\alpha - 3\beta) \sum_\mu C_\mu^4 +  \\
&&+ \left(12\beta-\gamma\frac{2+\eps}{4-2\eps}\right) \tr_R C^4 + 2 \, \beta \, \tr_R [C_\mu,C_\nu]^2\;. \nn
\eea
In appendix \ref{latticesums} it is shown that even though $\alpha, \beta$ and $\gamma$ are all divergent the combinations
appearing above for the terms that were not present at tree level, namely $C^2$, $C^4$ and $\sum_\mu C_\mu^4$, are all finite.
Only the coefficient of $[C_\mu,C_\nu]^2$ is divergent.

Now the full effective action (\ref{seff}) is easily written down using (\ref{abc}) in the adjoint representation together with (\ref{egy}) and 
in representation $R$ together with (\ref{ketto}). The traces of the product of two Lie algebra elements 
in different representations can be all converted to the fundamental
representation using the trace normalization factors $T(R)$ via $\tr_R( \cdot\;\cdot ) = 2\,T(R)\, \tr( \cdot\;\cdot )$. Let us first collect the
terms proportional to $\tr[C_\mu,C_\nu]^2$ which is the only divergent term. Using $T(ad) = N$ and the poles of
$\beta$ and $\gamma$ from appendix \ref{latticesums} we obtain,
\bea
\left.S_{eff}(C)\right|_{div} =  - \frac{(2\pi)^4}{2} \left( \frac{1}{g_0^2(\mu L)^{2\eps}} - \frac{\frac{11}{3}N - \frac{4}{3}T(R)N_f}{16\pi^2\eps} +
{\rm finite}  \right) \tr[C_\mu,C_\nu]^2  
\eea
Clearly, by introducing the renormalized coupling $g_R(\mu)$ of the MS scheme,
\bea
\frac{1}{g_R^2(\mu)} = \frac{1}{g_0^2(\mu L)^{2\eps}} - \frac{\frac{11}{3}N - \frac{4}{3}T(R)N_f}{16\pi^2\eps}\;,
\eea
in place of the bare coupling $g_0$ a finite effective action is obtained. Going from MS to $\msbar$ scheme only modifies the finite
terms.

Up until this point the momentum sums corresponding to the fermions were computed with periodic boundary conditions, however we are
interested in fermions that are anti-periodic in at least one direction. Instead of the coefficients $\alpha, \beta$ and $\gamma$
we should have considered $\alpha_\mu(k), \beta_{\mu\nu}(k)$ and $\gamma(k)$,
\bea
\alpha_\mu(k) &=& \sum_n \frac{(n_\mu-k_\mu)^4}{(n-k)^8} \nn \\
\beta_{\mu\nu}(k) &=& \sum_n \frac{(n_\mu-k_\mu)^2 (n_\nu-k_\nu)^2}{(n-k)^8} \\
\gamma(k) &=& \sum_n \frac{1}{(n-k)^4}\;, \nn
\eea
where $k_\mu \neq 0$ determines the boundary conditions.
However, it is easy to see that the differences $\alpha_\mu(k) - \alpha$, $\beta_{\mu\nu}(k) - \beta$ and $\gamma(k) - \gamma$ are
all finite. This is expected because UV divergences are insensitive to boundary conditions. Hence once the UV divergences are
canceled only the finite terms can be effected by the change of boundary conditions.

Summarizing this section, a finite effective action is obtained for the gauge zero modes of the form,
\bea
\label{fullseff}
S_{eff}(C) &=& - \frac{(2\pi)^4}{2g_R^2(\mu)} \tr [C_\mu,C_\nu]^2 + \\
&& + u_1 \tr C^2 + u_2 \tr_R C^4 + u_3 \tr_{ad} C^4 + \nn \\
&& + u_4 \sum_\mu \tr_R C_\mu^4 + u_5 \sum_\mu \tr_{ad} C_\mu^4\;, \nn
\eea
where the finite expressions $u_1, \ldots, u_5$ depend on $N$, $N_f$, $R$ and the boundary condition for the fermions.
These are all known although in a bit cumbersome form. Their values will not be important for what follows, the only property we need
is their finiteness. From now on we set $\mu = 1/L$.

\section{Yang-Mills gradient flow on $T^4$}
\label{ymgrad}

Now that a finite action is obtained for the gauge zero modes $C_\mu$ let us turn to our observable of interest, the field
strength squared $E(t)$ at positive flow time (\ref{e}). It will be evaluated by treating the gauge non-zero modes
in perturbation theory and the zero mode $C_\mu$ exactly, similarly to the effective action. Let us first write down the 
Yang-Mills gradient flow,
\bea
\frac{dA_\mu}{dt} = D_\nu F_{\nu\mu}\,.
\eea
Using the decomposition (\ref{abq}) we obtain a coupled flow for the zero and non-zero modes. After taking into account gauge
fixing and dropping terms higher order in $Q_\mu$ we arrive at,
\bea
\frac{dB_\mu}{dt} &=& [B_\nu,[B_\nu,B_\mu]]  \\
\frac{dQ_\mu}{dt} &=& \left( D_\rho(B)^2 \delta_{\mu\nu} + 2 [B_\mu,B_\nu] \right) Q_\nu\;. \nn
\eea
Since we are interested in a perturbative expansion let us rescale $Q_\mu \to g_R Q_\mu$. The consistent rescaling of the zero
mode is $B_\mu \to g_R^{1/2} B_\mu$. After the rescaling the gradient flow becomes
\bea
\frac{dB_\mu}{dt} &=& g_R [B_\nu,[B_\nu,B_\mu]]  \\
\frac{dQ_\mu}{dt} &=& \Delta Q_\mu + O(g_R^{1/2}) \;. \nn
\eea
Clearly, to leading order in the coupling the zero mode is constant $B_\mu(t) = B_\mu$ and the solution for the non-zero mode is
\bea
Q_\mu(t) = e^{t\Delta} Q_\mu(0)\;.
\eea
In the path integral one integrates over the fields at $t=0$, i.e. $Q_\mu(0)$ and $B_\mu$.

The rescaling also effects the observable $E(t)$ and keeping the leading order term only we obtain,
\bea
\label{et}
E(t) = - \frac{g_R^2}{2} \tr [B_\mu,B_\nu]^2 + \frac{g_R^2}{2} \tr\, Q_\mu e^{2t\Delta} \left( \Delta \delta_{\mu\nu} - \partial_\mu \partial_\nu \right) Q_\nu
\eea
where $Q_\mu$ now stands for $Q_\mu(0)$ for the sake of brevity.

Let us evaluate $\langle E(t) \rangle$ by first integrating over $Q_\mu$ while $B_\mu$ is kept fixed. 
The first term in (\ref{et}) is independent of $Q_\mu$ and the second term is quadratic, leading to
\bea
\label{eee}
\left.\langle E(t) \rangle\right._B = - \frac{g_R^2}{2} \tr [B_\mu,B_\nu]^2 + \frac{g_R^2}{2L^4} \tr\, e^{2t\Delta} \left( \Delta \delta_{\mu\nu} - \partial_\mu
\partial_\nu \right) \frac{\delta_{\mu\nu}}{\Delta} 
\eea
where only the leading order propagator is taken into account from the action. The integral and trace in the second term is given by,
\bea
\label{bbb}
&&3(N^2-1)\mysum e^{ - \pi^2 n^2 8t/L^2 } = \\
&&= 3(N^2-1)\left( \vartheta^4( e^{ -\pi^2 c^2 } ) - 1\right) = 
3(N^2-1)\left(\frac{1}{\pi^2c^4}\vartheta^4\left(e^{-1/c^2}\right) - 1\right)\;, \nn
\eea
using equation (\ref{xx}) from the appendix and where the ratio $c = \sqrt{8t}/L$ was introduced. The factor 3 comes from the
trace over the Euclidean indices and the factor $N^2-1$ comes from the gauge trace.

\begin{figure}
\begin{center}
\includegraphics[width=10cm]{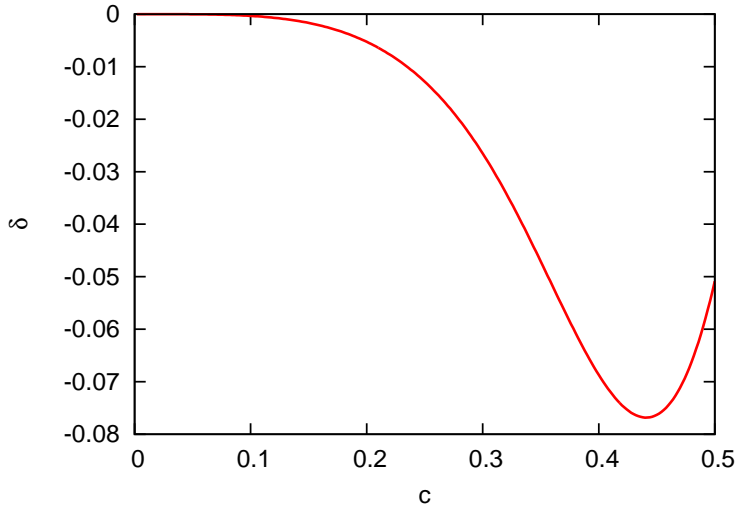}
\end{center}
\caption{Finite volume correction factor $\delta(c)$.}
\label{delta}
\end{figure}

Let us now integrate over $B_\mu$ using the effective action (\ref{fullseff}). One needs to keep the tree level part only, all
further terms are higher order in $g_R$. The second term in (\ref{eee}) is independent of $B_\mu$ while
for the first term we need the matrix integral
\bea
\label{zb}
-\frac{ \int dB \frac{1}{2} \tr[B_\mu,B_\nu]^2 \exp\left( \frac{L^4}{2} \tr [B_\mu,B_\nu]^2 \right)}{ \int dB \exp\left(
\frac{L^4}{2} \tr [B_\mu,B_\nu]^2 \right) } = \frac{N^2 - 1}{L^4}\;.
\eea
Even though the integral is quartic it can easily be done with the result $N^2 - 1$ essentially determined by the dimensionality
of the integral. Combining (\ref{eee}), (\ref{bbb}) and (\ref{zb}) we obtain,
\bea
\label{fin}
\langle t^2 E(t) \rangle = g_R^2\frac{3(N^2-1)}{128\pi^2}\left( 1 + \vartheta^4\left(e^{-1/c^2}\right) - 1 - \frac{c^4\pi^2}{3} \right)
\eea
which is the advertised final result (\ref{run}). The finite volume correction term $\delta(c)$ is plotted on figure \ref{delta} as a
function of the ratio $c$. As can be seen the correction never reaches $10\%$ for $0\leq c \leq 1/2$.

\section{Running coupling}
\label{runningcoupling}

The result (\ref{fin}) can be used to define a non-perturbative running coupling scheme in which the running scale is $\mu = 1/L$. As one changes
the scale one keeps $c$ fixed. Then the scheme is defined by the coupling constant
\bea
\label{gsq}
g_c^2(L) = \frac{ 128 \pi^2 \langle t^2 E(t) \rangle }{3(N^2-1)(1+\delta(c))}
\eea
where now the expectation value on the right hand side is understood non-perturbatively. The results from the preceding sections
ensure that the above defined coupling for small $L$ will run according to the universal 1-loop $\beta$-function. Different
choices for $c$ correspond to different schemes.

A note is in order about the 2-loop $\beta$-function. As is well known both the 1 and 2 loop coefficients are universal under a
scheme change of the type ${\tilde g} = g ( 1 + O(g^2) )$ where the expansion on the right hand side only contains even
powers of the coupling. However if one allows scheme changes of the type ${\tilde g} = g ( 1 + O(g) )$ where the expansion contains
both even and odd powers then only the 1-loop coefficient remains scheme independent. Our scheme is related to the
$\msbar$ scheme by such an expansion since it is easy to see that both even and odd powers of the coupling will appear as subleading
terms to the leading result (\ref{run}) but fractional powers will not. Our scheme is nevertheless well-defined and has
for instance the property that if a theory has an infrared fixed point in one scheme it will have a fixed point in our
scheme as well.

In order for the system to be controlled by a single scale $L$ the bare fermion mass was set to zero in the preceding sections.
The spectrum of the Dirac operator nevertheless has a gap $\sim 1/L$ due to the non-trivial boundary conditions for the fermions.

\section{Numerical results}
\label{numericalresults}

\begin{figure}
\begin{center}
\includegraphics[width=13cm]{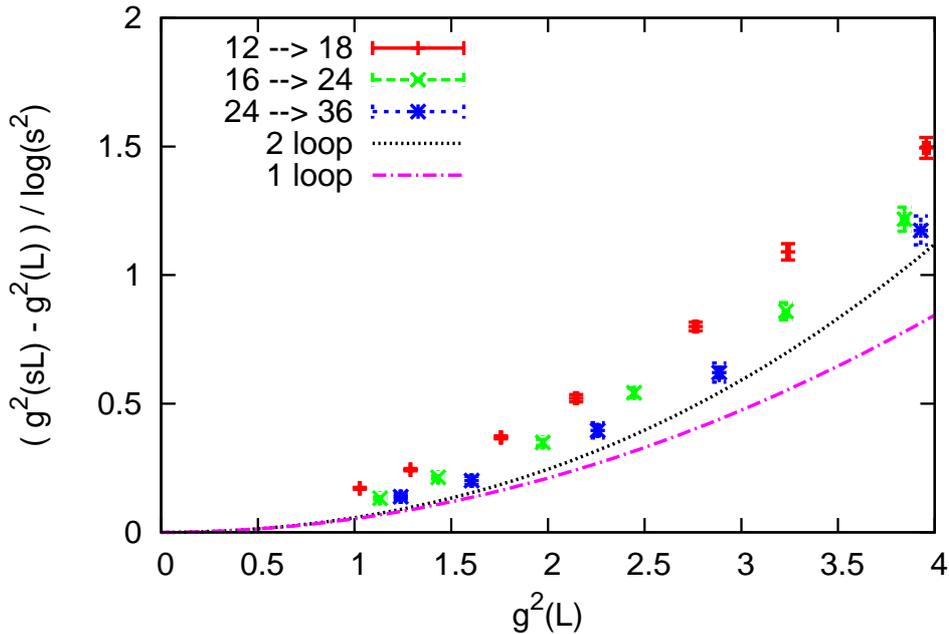}
\end{center}
\caption{Discrete $\beta$-function of $SU(3)$ gauge theory coupled to $N_f = 4$ flavors of massless fundamental fermions for a
scale change of $s=3/2$. The results at 3 lattice spacings are shown together with
the continuum 1 and 2-loop result from (\ref{dbeta}) for comparison.}
\label{beta}
\end{figure}

We have tested the new running coupling scheme in $SU(3)$ gauge theory coupled to $N_f = 4$ massless fundamental fermions. 
The Schr\"odinger functional analysis of the same model can be found in \cite{Tekin:2010mm, PerezRubio:2010ke}.
The fermion action was the 4-step stout improved \cite{Morningstar:2003gk} staggered action with smearing parameter $\varrho = 0.12$. Since the number of flavors is
a multiple of four no rooting was necessary. For the gauge sector tree level improved Symanzik action \cite{Symanzik:1983dc, Luscher:1984xn}
was used. The hybrid Monte Carlo
algorithm \cite{Duane:1987de} was used together with multiple time scales \cite{Sexton:1992nu} and Omelyan integrator \cite{Takaishi:2005tz}.

The observable $E(t)$ and the flow itself can be discretized in a number of ways. Both the discretization in
\cite{Luscher:2010iy} and also the tree level improved Symanzik discretization of \cite{Borsanyi:2012zs} was measured. We have
found that the latter displays better scaling as expected hence in the following only the results from the Symanzik discretization
will be presented. The bare quark mass was set to zero and anti-periodic
boundary conditions were used for the fermions in all four directions. As mentioned in the previous section this leads to a
gap $\sim 1/L$ in the spectrum of the Dirac operator. The gauge field was periodic in all directions.

The choice of $0\leq c \leq 1/2$ is limited by the observations that a small $c$ leads to large cut-off effects while large $c$ leads to
large statistical errors. We found that $c=0.3$ is a convenient choice and from here on will drop the index $c$ or $R$ on the
renormalized coupling $g^2$.

The discrete version of the $\beta$-function, or step
scaling function, was computed for a scale change of $s=3/2$. Three lattice spacings are used corresponding to
$12^4 \to 18^4$, $16^4 \to 24^4$ and $24^4 \to 36^4$. Then the discrete $\beta$-function
\bea
\frac{ g^2(sL) - g^2(L) } { \log(s^2) }
\eea
can be calculated as a function of $g^2(L)$. Holding $L$ fixed in physical units the continuum limit corresponds to $L/a \to\infty$.

\begin{table}
\begin{center}
\begin{tabular}{|l||c|c|c|c|c|c|c|c|}
\hline
$L/a \;\;\beta$ &	4.25	 &	4.50	 &	4.75	 &	5.00	 &	5.50	 &	6.00	 &	7.00	 &	8.00     \\
\hline
\hline
12              & 5.08(1)	 & 3.96(1)	 & 3.241(6)	 & 2.764(9)	 & 2.146(8)	 & 1.757(3)	 & 1.289(2)	 & 1.027(2)	 \\
\hline
16              & 6.41(3)	 & 4.79(2)	 & 3.84(2)	 & 3.23(1)	 & 2.446(6)	 & 1.974(5)	 & 1.432(2)	 & 1.132(3)	 \\
\hline
18              & 7.05(3)	 & 5.17(3)	 & 4.13(3)	 & 3.41(1)	 & 2.569(9)	 & 2.056(3)	 & 1.486(3)	 & 1.166(2)	 \\
\hline
24              &    	         & 6.34(4)	 & 4.83(3)	 & 3.93(2)	 & 2.89(1)	 & 2.257(9)	 & 1.605(5)	 & 1.239(4)	 \\
\hline
36              &   	         &      	 & 6.19(5)	 & 4.88(4)	 & 3.39(3)	 & 2.58(2)	 & 1.77(1)	 & 1.352(8)	 \\
\hline
\end{tabular}
\end{center}
\caption{Measured renormalized couplings $g_c^2(L)$ from (\ref{gsq}) at $c=0.3$ and given bare couplings $\beta$ and lattice
volumes $L/a$.}
\label{res}
\end{table}

The numerical results can be compared with the perturbative $\beta$-function for small renormalized couplings. The 2-loop $\beta$-function
is given by
\bea
L^2 \frac{dg^2}{dL^2} = b_1 \frac{g^4}{16\pi^2} + b_2 \frac{g^6}{(16\pi^2)^2}\,,\qquad b_1 = \frac{25}{3}\,,\qquad b_2 = \frac{154}{3}\;.
\eea
The discrete $\beta$-function up to 2 loops for a finite scale change $s$ is then
\bea
\label{dbeta}
\frac{g^2(sL)-g^2(L)}{\log(s^2)} = b_1 \frac{g^4(L)}{16\pi^2} + \left( b_1^2 \log(s^2) + b_2 \right) \frac{g^6(L)}{(16\pi^2)^2}\;,
\eea
which will be used for comparison although the zero mode of our finite volume scheme will introduce modifications which have not
yet been calculated.

The measured results for the renormalized coupling at each bare coupling and lattice volume are tabulated in table \ref{res}. 
At the volumes $12^4$, $16^4$, $18^4$, $24^4$ and $36^4$ the number of equilibrium trajectories were $10000$, $10000$, $10000$,
$8000$ and $4000$, respectively and every $10^{th}$ configuration was used for measurements. Auto correlation times were also
measured and are around $10 - 30$, $10 - 40$, $10 - 70$, $30 - 100$, $30 - 100$ for the five volumes, respectively.
The lower auto correlation times in the indicated intervals correspond to larger $\beta$ and the higher ones to smaller $\beta$.

The discrete $\beta$-function obtained from the data is shown on figure \ref{beta}. The continuum extrapolation can be performed in (at least) two different ways. In the first method a cubic spline interpolation
is done at fixed $L/a \to sL/a$ for $( g^2(sL) - g^2(L) ) / \log(s^2)$ as a function of $g^2(L)$.
Then the resulting three curves together with their errors are used
for the continuum limit at each fixed $g^2(L)$. The continuum extrapolation is linear in $a^2/L^2$ since both the action and the
observable only contain $O(a^2)$ corrections.
This latter step is repeated for each value of $g^2(L)$. 

\begin{figure}
\begin{center}
\includegraphics[width=13cm]{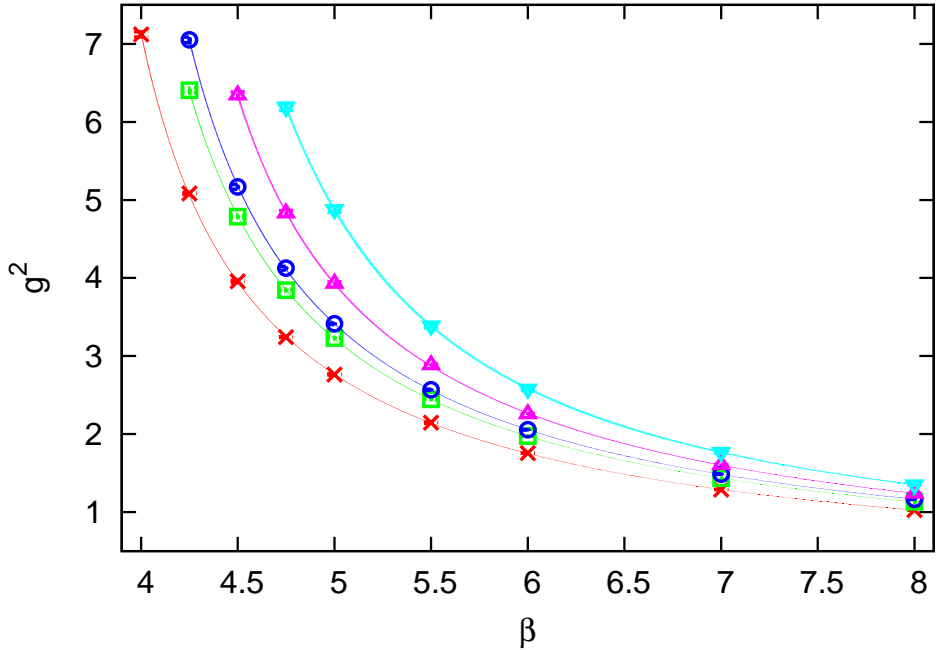}
\end{center}
\caption{Parametrization of the curves $g^2(\beta)$ at fixed lattice volumes using the expression (\ref{p}). Red: $12^4$, green:
$16^4$, dark blue: $18^4$, magenta: $24^4$, light blue: $36^4$.}
\label{para}
\end{figure}

In the second method, similarly to \cite{Tekin:2010mm}, the dependence of $g^2(\beta)$ on $\beta$ at fixed $L/a$ is parametrized by the
expression
\bea
\label{p}
\frac{\beta}{6} - \frac{1}{g^2(\beta)} = \sum_{m=0}^3 c_m \left( \frac{6}{\beta} \right)^m\;,
\eea
and the coefficients $c_m$ are fixed by fitting to the measured values. The $\chi^2/dof$ values from the fits for the five volumes are
$1.59$, $0.39$, $0.45$, $1.11$ and $0.08$, respectively from $12^4$ to $36^4$. The fitted curves together with the data are shown on figure \ref{para}.
Since the parametrization is linear in the coefficients $c_m$ the error on the fitted curve can be computed in a straightforward
manner. Then $g^2(L)$ together with the discrete $\beta$-function $( g^2(sL) - g^2(L) ) / \log(s^2)$ and its error can be obtained for any
$\beta$ for all three lattice spacings corresponding to $12^4 \to 18^4$, $16^4 \to 24^4$ and $24^4 \to 36^4$. From here the
procedure is identical to the previous method; at fixed $g^2(L)$ the three discrete $\beta$-function values are extrapolated to the
continuum assuming $O(a^2/L^2)$ corrections.

\begin{figure}
\begin{center}
\includegraphics[width=7.5cm]{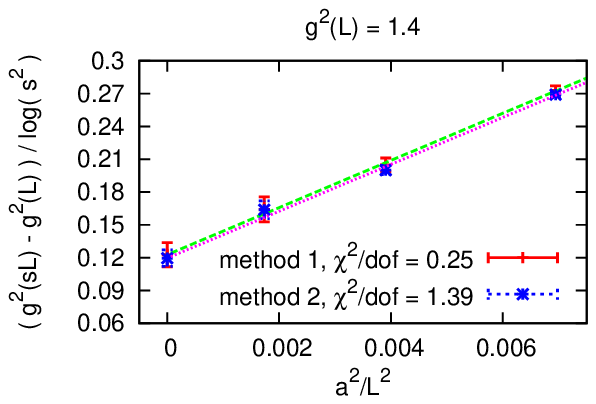} \includegraphics[width=7.5cm]{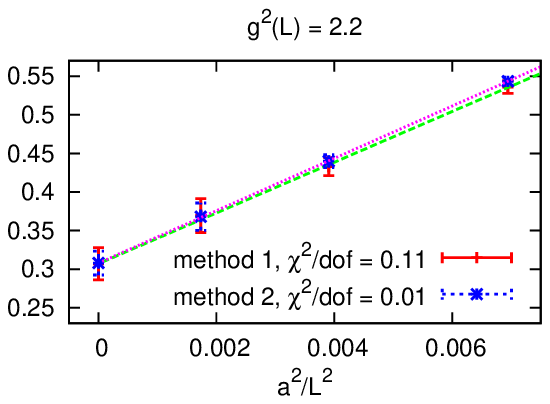} 
\includegraphics[width=7.5cm]{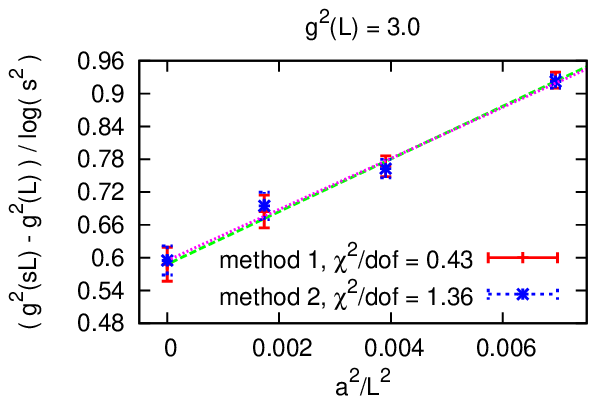} \includegraphics[width=7.5cm]{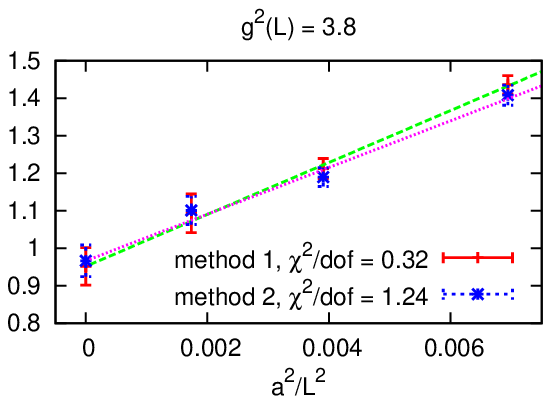}
\end{center}
\caption{Continuum extrapolations of the discrete $\beta$-function for four selected $g^2(L)$ values $1.4$, $2.2$, $3.0$ and
$3.8$. Both methods are shown together with the $\chi^2/dof$ values of the fits.}
\label{extrap}
\end{figure}

The continuum extrapolation is shown on figure \ref{extrap} for both methods and for 
four representative values of $g^2(L)$, $1.4$, $2.2$, $3.0$ and $3.8$ together with the $\chi^2/dof$ values of the fits. The continuum results agree
nicely between the two methods.

It is reassuring to note that the continuum extrapolations from the two methods yield continuum results that agree with each other within
error showing the robustness of the procedures. Also the continuum result is quite insensitive to the order of the
polynomial used in (\ref{p}) or other details of the fitting procedures.

\begin{figure}
\begin{center}
\includegraphics[width=13cm]{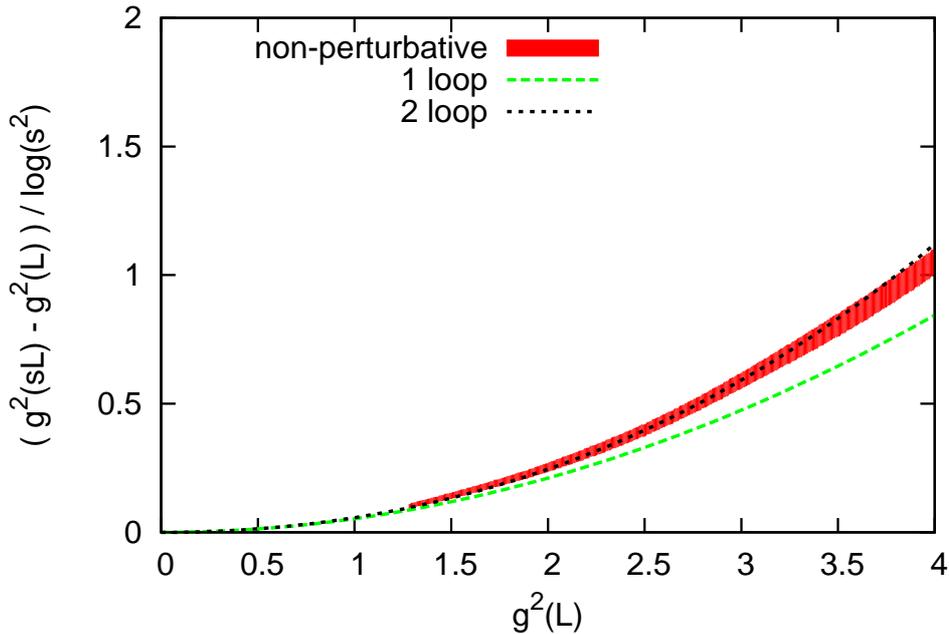}
\end{center}
\caption{Discrete $\beta$-function of $SU(3)$ gauge theory coupled to $N_f = 4$ flavors of massless fundamental fermions for a
scale change of $s=3/2$. The continuum extrapolated result from method 1 (see text for details) is 
shown together with the 1 and 2-loop results from (\ref{dbeta}) for comparison.}
\label{cont}
\end{figure}

The final continuum extrapolated result agrees approximately with the 2-loop perturbative expression (\ref{dbeta}) as shown on
figure \ref{cont} (only
the final result from the first method is shown, but the second one gives a result which agrees with it within errors in the entire
$g^2(L)$ range). As noted in section \ref{runningcoupling} our scheme is related to the $\msbar$ scheme via 
$g_c^2 = g_{\msbar}^2 ( 1 + a_1(c) g_{\msbar} + \ldots )$ 
where $a_1(c)$ is non-zero leaving only the first $\beta$-function coefficient scheme independent. 
It can be shown from the measured gradient flow at $c=0.2$ that the discrete $\beta$-function in 
figure \ref{cont} is not sensitive to the volume beyond
the leading $\delta(c)$ correction factor. This explains the approximate agreement with the 2-loop universal $\beta$-function keeping
contributions from $a_1(c)$ undetectable within errors.

\section{Conclusion and outlook}
\label{conclusion}

The Yang-Mills gradient flow -- or as implemented on the lattice, the Wilson or Symanzik flow -- is a promising tool for lattice gauge theory.
In order to use it for a running coupling scheme where the running scale is the size of the finite 4-dimensional box one needs to
compute the flow in perturbation theory at finite volume. In this work the necessary calculations were done in dimensional
regularization and the $\msbar$ scheme. The fact that the volume is finite necessitates the separation of the gauge Fourier modes
into zero and non-zero modes. The non-zero modes can be treated in 1-loop perturbation theory while the non-trivially interacting
zero modes need to be treated exactly. The result at leading order contains both algebraically and exponentially suppressed finite
volume correction terms relative to the infinite volume result.

The flow was then used to define a new scheme for the renormalized running coupling, agreeing with all other schemes at small
couplings, as it should. The new non-perturbatively well-defined scheme is actually a one
parameter family of couplings, all behaving universally for small values. The free parameter can be optimized for various targets
such as small cut-off effects and/or small statistical errors. Further advantages of the scheme is the fact that the necessary
observable can be evaluated in a Monte Carlo simulation at small computational cost relative to the HMC algorithm itself because
only gluonic observables are needed. In addition no extrapolation is needed for large Euclidean times.

The flow in infinite volume, as noted in \cite{Luscher:2010iy}, can also be used for a running coupling definition. In this setup
the coupling runs with the scale $\mu = 1/\sqrt{8t}$. Lattice implementation of this running over many orders of magnitudes 
requires additional control to keep finite volume effects small. The
scheme presented in this work circumvents this problem because the running scale is the volume itself, similarly to the
Schr\"odinger functional method \cite{Luscher:1992an}.

It would be very interesting to calculate further terms in the expansion (\ref{run}) as well as the leading cut-off effects to it.
Also, possible tunneling events at small renormalized coupling need to be investigated
in the future but in the numerical simulations so far we have not observed any for the measured observable.

\acknowledgments

This work was supported by the EU Framework Programme 7 grant (FP7/2007-2013)/ERC No 208740, by the Deutsche Forschungsgemeinschaft
grants FO 502/2 and SFB-TR 55, by the NSF under grants 0704171 and 0970137 and by the DOE under grants DOE-FG03-97ER40546 and
DOE-FG02-97ER25308. Computations were performed on the GPU clusters at the University of Wuppertal, Germany and at the Eotvos
University in Budapest, Hungary using the CUDA port of the code \cite{Egri:2006zm}. Kalman Szabo and Sandor Katz are
gratefully acknowledged for code development. KH wishes to thank the Institute for Theoretical Physics and the Albert Einstein
Center for Fundamental Physics at Bern University for their support. DN would like to thank Kalman Szabo for
suggesting to look into \cite{Luscher:2010iy} for possible applications and also Pierre van Baal for very useful discussions and
correspondence during the past 10 years on the small volume expansion of gauge theories. 

\appendix
\section{Conventions}
\label{conventions}

The gauge field is taken to be anti-hermitian and the covariant derivative and field strength tensor are given by
\bea
D_\mu &=& \partial_\mu + A_\mu \nn \\
F_{\mu\nu} &=& \partial_\mu A_\nu - \partial_\nu A_\mu + [ A_\mu, A_\nu ]
\eea
The $SU(N)$ trace is then negative definite. The notation $D_\mu(B) = \partial_\mu + B_\mu$ is also used and if it acts on gauge
fields it is understood to be in the adjoint representation. When acting on fermions it acts in the representation $R$.

The Dirac operator $\slashed{D} = \gamma_\mu D_\mu$ is also implicitly acting in representation $R$. For the commutators of gamma
matrices $\sigma_{\mu\nu} = \half ( \gamma_\mu \gamma_\nu - \gamma_\nu \gamma_\mu )$ is used.

\section{Evaluation of momentum sums}
\label{latticesums}

The Jacobi elliptic function is used in the text,
\bea
\vartheta(q) = \sum_{n=-\infty}^{\infty} q^{n^2}\;,
\eea
which has the property
\bea
\label{xx}
\vartheta\left(e^{-\pi/t}\right) = \sqrt{t} \vartheta\left( e^{-\pi t} \right)\;.
\eea
Let us introduce $h(t) = \vartheta(\exp(-t\pi))$. Then we have, using property (\ref{xx}),
\bea
\mysum \frac{1}{n^{2s}} &=& \frac{\pi^s}{\Gamma(s)} \int_0^\infty \,dt\, t^{s-1}\, \left( h^d(t) - 1 \right) = \nn \\
&=& \frac{\pi^s}{\Gamma(s)} \left( \frac{2d}{2s(2s-d)} + \int_1^\infty\, dt \,\left( t^{d/2-1-s} + t^{s-1} \right) \left( h^d(t) - 1 \right) \right)\,,
\eea
which leads to the finite value $\delta = -5.545177$ for $s=1$ and $d=4$. For the other two constants we similarly have,
\bea
\mysum \frac{n_1^4}{n^{2s}} &=& \frac{\pi^{s/2}}{\Gamma(s/2+2)} \int_0^\infty\, dt\, t^{s/2+1} h^{\prime\prime}(t) h^{d-1}(t) \nn \\
\mysum \frac{n_1^2n_2^2}{n^{2s}} &=& \frac{\pi^{s/2}}{\Gamma(s/2+2)} \int_0^\infty\, dt\, t^{s/2+1} h^{\prime\,2}(t) h^{d-2}(t)\,,
\eea
and using again property (\ref{xx}) we obtain for $s=4$ and $d=4-2\eps$,
\bea
\alpha = \frac{\pi^2}{8\eps} + 1.622255 + O(\eps)\,,\qquad \beta = \frac{\pi^2}{24\eps} - 0.063112 + O(\eps)\,.
\eea
Clearly, $\alpha - 3\beta$ is finite. The last remaining constant $\gamma$ is not independent from the rest, 
we have $\gamma = d \alpha + d (d-1)\beta$, leading to
\bea
\gamma = \frac{\pi^2}{\eps} - 2.492991 + O(\eps)\,.
\eea

\end{document}